 \documentclass[final,5p,times,twocolumn,authoryear]{elsarticle}

\usepackage{amssymb}
\usepackage{lipsum}

\usepackage{url}
\usepackage{amsmath}

\def\lum{$\rm erg~s^{-1}$}

\def\arcsec{\hbox{$^{\prime\prime}$}}

\def\srge{SRGe~J021932}

\newcommand{\ero}{eROSITA}	

\newcommand{\srg}{{\it SRG}}
\newcommand{\xmm}{{\it XMM-Newton}}
\newcommand{\gaia}{{\it Gaia}}

\journal{High Energy Astrophysics}

\begin{document}

\begin{frontmatter}



\title{Superflare on a rapidly-rotating solar-type star captured in X-rays}
\author[first,second]{Andrey Mukhin\corref{cor1}}
\cortext[cor1]{Corresponding author}
\ead{amukhin@cosmos.ru}

\author[first]{Roman Krivonos}
\author[third,forth]{Ilfan Bikmaev}
\author[third]{Mark Gorbachev}
\author[third]{Irek Khamitov}
\author[first]{Sergey Sazonov}
\author[first,fifth]{Marat Gilfanov}
\author[first,fifth,sixth]{Rashid Sunyaev}

\affiliation[first]{organization={Space Research Institute (IKI), Russian Academy of Sciences},
            city={Moscow},
            postcode={117997}, 
            country={Russia}}
\affiliation[second]{organization={Moscow Institute of Physics and Technology (National Research University)},
            addressline={}, 
            city={Moscow},
            postcode={117303}, 
            country={Russia}}
\affiliation[third]{organization={Kazan Federal University},
            addressline={Kremlevskaya Str.18}, 
            city={Kazan},
            postcode={420008}, 
            country={Russia}}
\affiliation[forth]{organization={Academy of Sciences of Tatarstan Republic},
            addressline={Baumana Str. 20}, 
            city={Kazan},
            postcode={420111}, 
            country={Russia}}
\affiliation[fifth]{organization={Max-Planck-Institut fur Astrophysik},
            addressline={Karl-Schwarzschild-Str. 1}, 
            city={Garching},
            postcode={D-85741}, 
            country={Germany}}
\affiliation[sixth]{organization={Institute for Advanced Study},
            addressline={1 Einstein Drive}, 
            city={Princeton},
            postcode={08540}, 
            state={New Jersey},
            country={USA}}

\begin{abstract}
In this work, we studied X-ray source SRGe~J021932.4$-$040154 ({\srge}), which we associated with a single X-ray active star of spectral class G2V$-$G4V and the rotational period $\rm P_{rot} < 9.3$ days. Additional analysis of TESS light-curves allowed for the rotational period estimation of $3.2 \pm 0.5$ days. {\srge} was observed with the {\srg}/{\ero} during eUDS survey in 2019 in a much dimmer state compared to the {\xmm} catalogue 4XMM-DR12. Detailed analysis revealed that the archival {\xmm} observations captured the source during a flaring event in 2017. The {\xmm} light curve demonstrates a strong flare described with the Gaussian rise and exponential decay, typical for stellar flares, characterized by timescales of ${\sim}400$~s and ${\sim}1300$~s, respectively. The spectral analysis of the quiescent state reveals ${\sim}10$~MK plasma at luminosity of $(1.4\pm0.4) \times 10^{29}$~\lum\ (0.3$-$4.5~keV). The spectrum of the flare is characterized by temperature of ${\sim}40$ MK and luminosity $(5.5\pm0.6)\times 10^{30}$~\lum. The total energy emitted during the flare ${\sim}1.7 \times 10^{34}$~erg exceeds the canonical threshold of $10^{33}$ erg, allowing us to classify the observed event as a superflare on a rapidly-rotating solar-type star. Additionally, we present the upper limit on the surface starspot area based on the brightness variations and consider the hypothesis of the object being a binary system with G-type and M-type stars, suggested by two independent estimations of radial velocity variations from APOGEE-2 and {\gaia}.
\end{abstract}




\end{frontmatter}




\section{Introduction}

Solar flares are powerful magnetic phenomena caused by the reconnection of magnetic lines in the solar corona with rapid release of magnetic energy \citep{Fletcher2011_overview}. This process leads to effective acceleration of electrons, that propagate downwards to the chromosphere, causing the heating of surrounding plasma up to millions degrees K. The pressure of heated plasma soon exceeds the pressure of chromosphere, leading to the expansion of the plasma and creation of a region of hot (with temperatures $>$ 10 MK) optically thin large volume of plasma, that eventually cools down to the usual stellar coronae temperature of 1-3 MK \citep{Benz2010_solarflares}. Different stages of the process have been observed in all wavelengths from radio to $\gamma-$rays, however, the emission of hot region and the solar corona is most prominent in X-rays.  

On the one hand, the solar flares allow us to study in details the stellar flares on G-dwarf stars, on the other hand, the investigation of stellar flares on distant G-dwarfs can reveal more extreme events, potentially being observed on the Sun. While the solar flares typically release $10^{29}-10^{32}$ erg in optical range at the timescale of several hours, some G-dwarfs were observed to demonstrate flaring events with energy release of $10^{33}-10^{36}$ erg, often called as ``superflares'' \citep{Schaefer2000_superflares, Maehara2012_superflares, Maehara2015_superflares, Tu2021_catalog}.

Superflares were already observed for a variety of stellar objects besides solar-type G-dwarfs: the RS CVn binaries (See, for example), young T Tauri stars, and UV Ceti-like red dwarfs \citep{Haisch1991_superflares}. Superflares in X-ray range, however, were observed only for several objects: Algol \citep{Favata1999_Algol}, AB~Doradus \citep{Maggio2000_ABDor, Didel2024_ABDor}, II~Pegasi \citep{Osten2007_IIPegasi}, EV~Lac \citep{Favata2000_EVLac, Osten2010_EVLac}, CC Eri \citep{Karmakar2017_CCEri}, UX~Arietis \citep{Franciosini2001_UXAri}, DG~CVn \citep{Fender2015_DGCvn}, SZ Psc \citep{Karmakar2023_SZPsc}, and some of them were even captured simultaneously in optical and X-ray range: AD Leonis \citep[XMM-Newton and TESS,][]{Stelzer2022_ADLeo}, ten K-M, one F9 and one G8 star in the Pleiades \citep[XMM-Newton and Kepler/K2,][]{Guarcello2019_Pleiades} and three more K-M stars using XMM-Newton and Kepler \citep{Kuznetsov2021_simsuperflares}. 

The majority of the hosts for these X-ray superflares turned out to be either M-stars or binary, or multiple systems with an M-component \citep{Karmakar2022_EQPeg}, however, there are examples of G-dwarf stars with supeflares both in optical and X-ray range with no observed companions or in wide binary systems: EK Draconis \citep{Namekata2024_EKDraconis}, $\rm \kappa^1~Ceti$ \citep{Namaguchi2023_k1Ceti} and several others \citep{Audard2000_superflares_in_late_type, Schaefer2000_superflares, Shibata2013_canoccur}.

In this study, we investigate the weak X-ray source SRGe~J021932.4$-$040154 (hereafter {\srge}) detected during the SRG/eROSITA \citep{Sunyaev2021_SRG, Predehl2021_erosita} X-ray survey of the UKIDSS Ultra Deep Survey field \citep[eUDS,][]{Krivonos2024_eUDS}, that was previously observed in much brighter X-ray state on January 1 2017 with XMM-Newton under the name 4XMM~J021932.$-$040153 \citep[4XMM-DR12 catalogue,][]{2020A&A...641A.136W}. {\srge} is associated with an optical G-dwarf star with no confirmed companions by several observatories. We reconstructed the XMM-Newton light-curve, which revealed the rapid increase in the flux followed by gradual decay, implying the possible flaring nature. By the observed total energy emission of $10^{34}$~erg and X-ray luminosity of ${\approx}10^{30}$\lum\ in 0.3$-$4.5~keV band, we classify this event as a superflare. We present XMM-Newton timing and spectral analysis of {\srge} in quiescent (non-flaring) state observed in 2016 and during the superflare captured in 2017, as well as SRG/eROSITA observations in 2019.

\section{Identification of {\srge}}

\cite{Krivonos2024_eUDS} report that X-ray emission of \srge\ is consistent with a point-like object (RA=2:19:32.523, Dec=-4:01:53.671, J2000), and its position is determined with $2.4''$ uncertainty (\textit{RADEC\_ERR}). We consider an association as likely if the optical position is within 1.5 $\times$ \textit{RADEC\_ERR}, which approximately
corresponds to 90\% position error. We performed search within this error radius in the catalogues of {\gaia} \citep{gaia2016_mission}, Transiting Exoplanet Survey Satellite  \citep[TESS,][]{Ricker2015_TESS} and APOGEE$-$2 of SDSS \citep{Planton2017_SDSS4}. The following counterparts have been found: {\gaia} DR3 2489887138146547456 \citep{Gaia2023_DR3} with 2.17{\arcsec} separation, TIC 332890423 from TESS Input Catalogue v8.0 (TIC-8) \citep{Stassun2019_TESSv8} with 2.54{\arcsec} separation and APOGEE ID 2M02193235$-$0401533 from APOGEE$-$2 SDSS DR17 \citep{APOGEE17_2022} with 2.59{\arcsec} separation. All three associations' coordinates are consistent with each other within 0.4{\arcsec}, thus we assume them be the observations of the same object 2MASS J02193235-0401533. All three catalogs report that the position of \srge\ is consistent with a star. 

\subsection{Optical properties of the star}
Table \ref{tab:cat_data} outlines the properties of all three associations, namely, effective temperature $\rm T_{eff}$, estimated distance, radial velocity RV, rotational velocity radial component $\rm v~sini$, surface gravity $\rm log~g$,$\rm [M/H]$ metallicity compared to Solar, radius, mass and bolometric luminosity of the object.

\begin{table*}
    \centering
    \begin{tabular}{llll}
        Property & {\gaia} DR3 & TIC-8 & APOGEE-2 DR17  \\
        \hline
        $\rm T_{eff}$, K & $\rm 5736 \pm 7$  & $\rm 5760 + 120$ & $\rm 5678 \pm 26$\\
        $\rm Distance$, pc & $347 \pm 5$ & $\rm 342 \pm 12$ & --\\
        RV, $\rm km~s^{-1}$ & $\rm 19.14 \pm 1.58$ & -- & $\rm 20.18 \pm 0.04$\\
        $\rm v~sini$, $\rm km~s^{-1}$ & -- & -- & 5.78 \\
        log g, $\rm cm~s^{-2}$ & $\rm 4.3296 \pm 0.008$ & $\rm 4.38 \pm 0.08$ & $\rm 4.18 \pm 0.02$ \\
        $\rm [M/H]$ & $-0.232 \pm 0.007$ & $0.004 \pm 0.014$ & $-0.024 \pm 0.005$ \\
        $\rm Radius, R_{\odot}$ & $\rm 1.04 \pm 0.01$ & $\rm 1.09 \pm 0.06$ & --\\
        Mass, $\rm M_{\odot}$ & $\rm 1.0 \pm 0.4$ & $\rm 1.0 \pm 0.1$ & --\\
        Luminosity, $\rm L_{\odot}$ & $\rm 1.19 \pm 0.04$ & $\rm 1.18 \pm 0.09$ & --\\
    \end{tabular}
    \caption{Properties of \srge\ associations with {\gaia} DR3 \citep{Gaia2023_DR3}, TESS Input Catalogue v8.0 \citep{Stassun2019_TESSv8} and APOGEE-2 SDSS DR17 \citep{APOGEE17_2022}.}
    \label{tab:cat_data}
\end{table*}

The effective temperature indicates the associated star to be G2$-$G4 spectral class. The surface gravity and the estimated radius are typical for the main sequence stars. Thus, we assume that {\srge} is an X-ray active yellow dwarf (G2V$-$G4V) at ${\approx}347$~pc from Earth.

\subsection{Binary system indications}
\label{sec:binary}

Although the 2MASS J02193235-0401533 is not reported as a binary system by any of the used catalogs, it should be noted, that there are two peculiar features present in APOGEE-2 SDSS DR17 and {\gaia} DR3 data, that can indicate a presence of a companion of the G-star.

APOGEE-2 SDSS DR17 reports the radial velocity to be determined with the uncertainties of 0.04 $\rm km~s^{-1}$, while the scatter of individual visits around average radial velocity is reported to be $0.53~\rm km~s{-1}$. The scatter of the individual visits much larger than typical uncertainties can be used as an indicator of radial velocity variations due to being in a binary system. Nevertheless, only 3 data points for the radial velocity estimation are currently present in APOGEE-2 SDSS DR17, and further observations are required to reliably detect variations.

{\gaia} DR3 estimations based on 12 transits show the radial velocity variation amplitude of $17.5~\rm km~s^{-1}$ despite the estimated uncertainties of only $1.6~\rm km~s^{-1}$. As well as for the SDSS, it can indicate radial velocity variability and the presence of the companion. As the radial velocity time series is not currently available, we estimate the probability of the radial velocity variations with the reported P-value for constancy of 0.64. While it cannot be used to reject neither the binary system hypothesis nor the single star hypothesis, we interpret the reported statistics to imply no significant radial velocity variations are observed.

Thus, in this work, we assume the {\srge} to be a single star, although further analysis is required to make a reliable conclusion.

\section{X-ray Observations}

\srge\ was observed with {\srg}/{\ero} in eUDS survey \citep{Krivonos2024_eUDS} conducted in August-September 2019 during the \srg\ Calibration and Performance Verification (Cal-PV) phase. The source attracted attention by its very dim flux compared to the XMM-Newton archival observations as reported in 4XMM-DR12 catalogue \citep{2020A&A...641A.136W}.

\subsection{SRG/eROSITA}

\begin{figure}
    \centering
    \includegraphics[width=\linewidth]{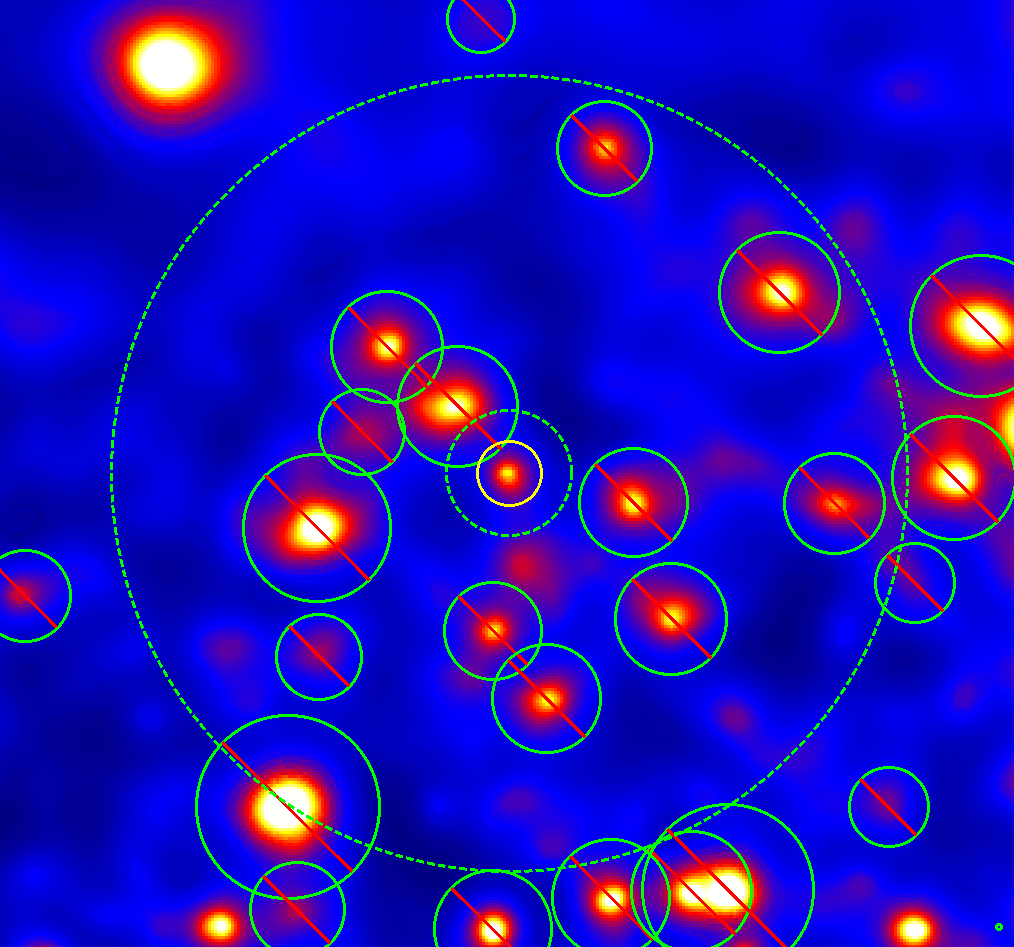}
    \caption{{\srg}/{\ero} 0.3--2.3~keV sky map of the region around {\srge} (in the center) in the units of cts~s$^{-1}$. The exposure-corrected image has been adaptively smoothed with the \texttt{dmimgadapt} task from \texttt{CIAO-4.15} \citep{ciao} using a Gaussian kernel. The image is shown in the square-root scale colour map,  \citep[``b'' in DS9 notation, see ][]{ds9} ranging from zero (black) to 0.001 and above (white). The X-ray spectrum of {\srge} has been extracted from yellow $R=43''$ circle. The background has been estimated in annulus region $R_{\rm min}=84''$ and $R_{\rm max}=530''$. The excluded regions are marked by a crossed out circles.}
    \label{fig:eromap}
\end{figure}

The processing of the {\srg}/{\ero} data was performed using the eSASS version 211201 software as described in \cite{Krivonos2024_eUDS}. We ran the \texttt{srctool} task in \texttt{AUTO} mode (when \texttt{srctool} generates source and background regions automatically) to extract spectral information between $0.2$ and $10$~keV of all eUDS sources, including {\srge}. The source regions are circles and the background regions are annuli, both centered on the position of the source. The size of the regions is selected taking the source counts, the level of the background, the best fitting source extent model radius, and other parameters into account\footnote{See the eSASS documentation at \url{https://erosita.mpe.mpg.de/edr}}. The result of the automatic generation of source and background regions is shown in Fig.~\ref{fig:eromap}. The source spectrum is extracted from the circle with $R=43''$ around the best-fit source position. The total count within this aperture is 44 over an exposure of 4.1~ks.  As seen from the figure, despite the crowded field around {\srge}, it is not affected by spatial confusion with nearby sources.

Additionally, {\srge} is present in the {\srg}/{\ero} catalog of X-ray sources in the eastern Galactic hemisphere based on the $\sim$4.4 {\srg} all sky surveys (December 2019 to February 2022) \citep{Khamitov2022_allsky}. The source was detected with ${\tt DET\_LIKE} = 10.2$ and a single optical {\gaia} DR3 catalog association inside $\rm R98 = 14.3{\arcsec}$. The X-ray luminosity of the source in 0.3$-$2.3 keV based on the {\srg}/{\ero} flux and {\gaia} parallax data is estimated as $\rm (2.2 \pm 0.8) \times 10^{29}~erg~s^{-1}$.

\subsection{XMM-Newton}

We use two observations of XMM-Newton \citep{Jansen2001_XMM}: ObsID 0785100401 on July 3 2016 and ObsID 0793580201 on January 1 2017 (Table~\ref{tab:xmm_obs}) conducted as a part of the XMM-SERVS survey \citep{Chen2018_SERVS}. Since MOS1 camera was not operating during both observations, we used data from PN and MOS2 cameras only. 

\begin{table}
    \centering
    \begin{tabular}{cccc}
        \hline
             ObsID & Date & Time (UTC) & T (ks)\\
        \hline
             0785100401 & 2016-07-03 & 03:44:16 - 09:50:56 & 22 \\
             0793580201 & 2017-01-01 & 07:02:08 - 09:32:08 & 9
    \end{tabular}
    \caption{{\xmm} observations of the {\srge} used in this work.}
    \label{tab:xmm_obs}
\end{table}

\begin{figure}
    \centering
    \includegraphics[width=\linewidth]{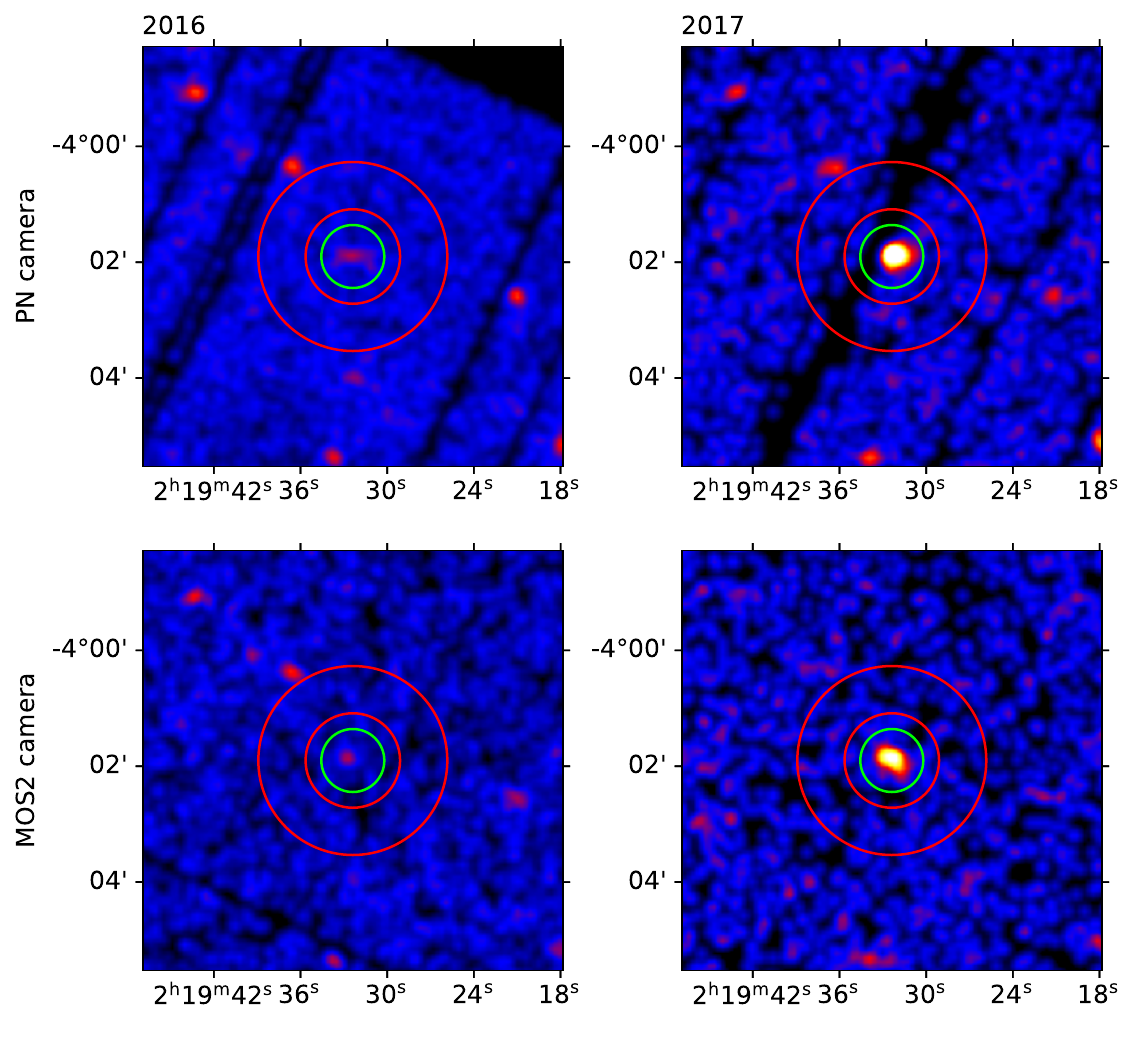}
    \caption{{\xmm} sky images around {\srge} in $\rm cts~s^{-1}$ with PN (top row, 0.2-15 keV) and MOS2 (bottom row, 0.2-10 keV) cameras for two {\xmm} observations from 2016 (left column) and 2017 (right column) observations as described in Table~\ref{tab:xmm_obs}. Images in each row are shown with the matching square-root scale colour maps ``b'' in DS9 notation, ranging from $10^{-6}$ to $10^{-3}$ for PN and from $10^{-6}$ to $5 \times 10^{-4}$ for MOS2. All images were smoothed with the Gaussian kernel ($\sigma = 1$~pix). The X-ray spectrum and light-curves for the SRGe J021932 have been extracted from the green circle region with $\rm R = 30 \arcsec$. Red annulus ($R_{\rm in} = 45 \arcsec$ and $R_{\rm out} = 90 \arcsec$) represents the region of background estimation.}
    \label{fig:xmm_img}
\end{figure}

The XMM-Newton observations were processed using the XMM-Newton Science Analysis System (SAS) v21.0.0 software. The observations were filtered for the flaring particle background, and the light-curves and spectra were generated using standard SAS Data Analysis Threads with default selection parameters for PN \texttt{(PATTERN<=4)} and MOS2 \texttt{(PATTERN<=12)}. We extracted light-curves and spectra from the source region described with  $R=30 \arcsec$ circle around {\srge} position and the background from the $45-90 \arcsec$ annulus as shown in Fig. \ref{fig:xmm_img}. The images are extracted with the same filtering parameters as the spectra and light-curves. The black streaks observed for the PN camera represent the edges of detectors \citep{Struder2001_PN_camera} and are taken into account in the standard XMM data processing tools.

\section{Analysis and results}

As seen from the sky images presented in Fig. \ref{fig:xmm_img}, {\xmm} captured {\srge} in two different states -- in 2016 it was observed being much dimmer and in 2017 being brighter for both cameras. Hereafter, we refer to the state observed during 2016 as a quiescent state and the state during 2017 as a flaring state.

\subsection{Light-curve analysis}
\label{sec:lc}

Assuming no significant variations within 20 ks exposure time (${\approx}5$~hours) during the quiescent state, we fit the 2016 light-curves of {\srge} with a constant. The resulting values for constant count rates are $\rm (1.2 \pm 0.5) \times 10^{-2}~cts~s^{-1}$ and $\rm (1.4 \pm 1.3) \times 10^{-3}~cts~s^{-1}$ for PN and MOS2, respectively.

For 2017 observation, the light-curves in both cameras show a rapid rise and gradual decay, which we fit as a sum of a constant count rate and an asymmetric peak with Gaussian rise and exponential decay. Thus, the model for fitting the light-curves of the flare is:

\begin{equation}
    \label{eq:time_model}
    C(t) = C_{\rm const} + A_{\rm flare} \times \begin{cases}
       {\rm exp}\left(-\left(\frac{t-t_{\rm peak}}{\tau_{\rm rise}}\right)^2\right),
       & \quad t \leq t_{\rm peak} \\
       {\rm exp}\left(-\frac{t-t_{\rm peak}}{\tau_{\rm decay}}\right),
       & \quad t > t_{\rm peak}
    \end{cases}
\end{equation}

where $C_{\rm const}$ is the constant count rate, $A_{\rm flare}$ is the amplitude of the flare, $t_{\rm peak}$ is the time position of the flare peak, $\tau_{\rm rise}$ and $\tau_{\rm decay}$ are the estimations rise and decay phase characteristic times, respectively.

For PN light-curve we fitted the data with the model with five free parameters ($C_{\rm const},~A_{\rm flare},~t_{\rm peak},~\tau_{\rm rise},~\tau_{\rm decay}$). MOS2 light-curve, due to poor statistics, was fitted with two free parameters $C_{\rm const},~A_{\rm flare}$ and other parameters fixed to the optimal values found for PN light-curve.

\begin{figure*}
    \centering
    \includegraphics[width=\linewidth]{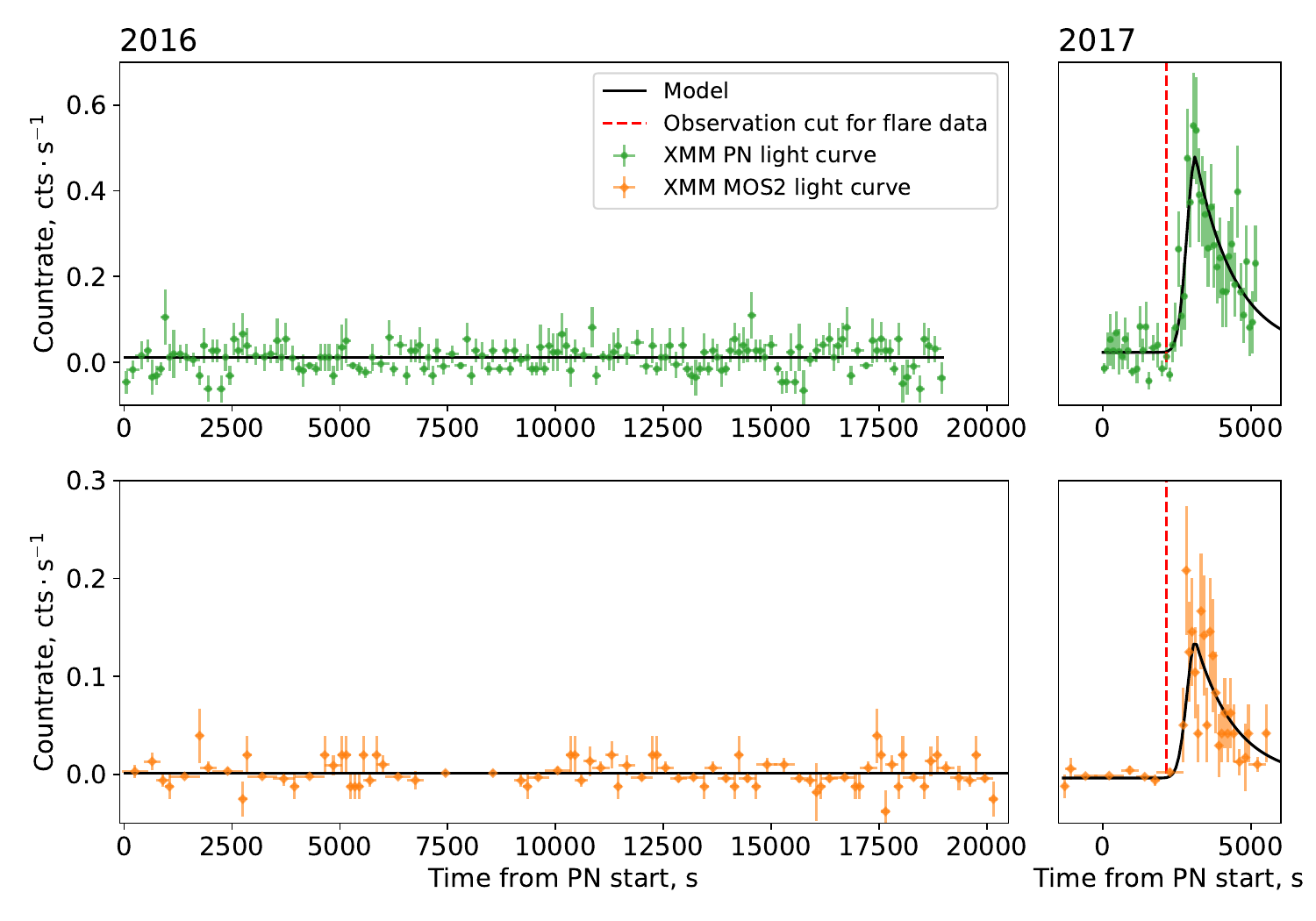}
    \caption{Light-curves of the source from XMM-Newton PN (top row) and MOS2 (bottom row) cameras for observations in the quiescent state (2016, left column) and during the flare (2017, right column). The black line represents the flare light-curve model described by the Eq. \ref{eq:time_model}. Red vertical dashed line shows the position of the estimated start of the flare.}
    \label{fig:timing}
\end{figure*}

The best-fit parameters of the flare light-curves with estimated 90\%  confidence intervals are presented in Table \ref{tab:timing}. The resulting model together with PN and MOS2 light-curves are shown in Fig. \ref{fig:timing}.

The estimated constant level count rates for 2017 observation are consistent with the constant count rates for 2016 observation. The rise and decay phases are characterized by $4.1_{-3.5}^{+1.4} \times 10^{2}~\rm s$ and $1.3_{-0.4}^{+0.6} \times 10^{3}~\rm s$.

We determined the nominal start of the flare ({$T_{\rm start} = 599646020~s$} or 2017 January 1 08:19:11 UTC) as the time when the best-fit flare model (Eq.~\ref{eq:time_model}) exceeds 1\% of the peak value $A_{\rm flare}$. In the following, we extract X-ray spectrum of the flare after this time.

\begin{table}
    \centering
    \begin{tabular}{cll}
                                    & PN & MOS2 \\
        \hline
        $C_{\rm const},~\rm cts~s^{-1}$ & $2.3^{+1.7}_{-1.2} \times 10^{-2}$ & $-0.1^{+0.3}_{-0.3} \times 10^{-2}$\\
        $A_{\rm flare},~\rm cts~s^{-1}$ & $4.6_{-0.7}^{+1.2} \times 10^{-1}$ & $1.3_{-0.3}^{+0.3} \times 10^{-1}$ \\
        $t_{\rm peak},~\rm s$           & $3.1_{-0.2}^{+0.1} \times 10^{3}$ & -- \\
        $\tau_{\rm rise},~\rm s$        & $4.1_{-3.5}^{+1.4} \times 10^{2}$ & -- \\
        $\tau_{\rm decay},~\rm s$       & $1.3_{-0.4}^{+0.6} \times 10^{3}$ & -- \\
        $\rm \chi_r/dof$            & 0.79/40 & 0.94/36 \\
    \end{tabular}

    \caption{Table of best-fit flare model parameters for XMM2 PN and MOS2 light-curves with 90\% confidence intervals calculated via bootstrapping. Empty cells in the second column correspond to the fixed parameters taken from PN model. Value of $\rm t_{peak}$ is calculated from the beginning of the PN camera observation.}
    \label{tab:timing}
\end{table}

\subsection{Spectral analysis}

For the spectral analysis, we have used a total of five spectra:

\begin{itemize}
    \item two, PN and MOS2, spectra in quiescent state (2016)
    \item two, PN and MOS2, spectra in flaring state (2017)
    \item a single {\srg}/{\ero} spectrum in quiescent state (2019)
\end{itemize}
The 2017 spectra are extracted within the time interval between  $T_{\rm start}$ (Sect.~\ref{sec:lc}) and the end of observation, resulting in a total exposure of 3050~s. All spectra were re-binned to contain at least 3 counts in a bin to perform fitting with Cash statistics \citep{Cash1979_stats}. For the spectral fitting, we used XSPEC \citep{Arnaud1996_XSPEC} software.

For the initial spectral analysis, we have fitted all 5 spectra separately with a simple power law model with fixed galactic neutral atomic hydrogen (NHI) column density $\rm 1.91\times10^{20}~cm^{-2}$ derived by \cite{HI4PI2016}. We use the complete column density estimation as the {\srge} is situated at a high galactic latitude ($b \sim -59^{\circ}$) and the distance of $343~\rm pc$. We then calculated the estimated flux in 0.3$-$4.5 keV range for every spectrum. The resulting values are presented in Table \ref{tab:powerlaw}.

\begin{table}
\centering
\begin{tabular}{lcll}
                & $\Gamma$ & Flux (0.3$-$4.5 keV) & cstat / bins\\
                &              & $\rm erg~s^{-1}~cm^{-2}$ & \\
    \hline
    2016 PN       & $1.6 \pm 0.4$ & $1.7_{-0.5}^{+0.6} \times 10^{-14}$ & 59.19 / 56\\
    2016 MOS2     & $2.4 \pm 0.7$ & $1.4_{-0.7}^{+0.8} \times 10^{-14}$ & 21.95 / 20\\
    2017 PN       & $1.7 \pm 0.2$ & $(4.3 \pm 0.6) \times 10^{-13}$ & 55.42 / 47\\
    2017 MOS2     & $1.6 \pm 0.3$ & $(3.2\pm0.6) \times 10^{-13}$ & 33.09 / 30 \\
    2019 {\ero} & $2.0 \pm 0.6$ & $1.5_{-0.6}^{+0.8} \times 10^{-14}$ & 39.59 / 31\\
\end{tabular}
\caption{Results of fitting of 5 X-ray {\srge} spectra with a simple powerlaw model with the fixed galactic neutral hydrogen column density absorption (\texttt{tbabs(powerlaw)}).}
\label{tab:powerlaw}
\end{table}

The X-ray spectrum of {\srge} observed by {\srg}/{\ero} in 2019 is consistent with PN/MOS2 spectra in 2016, while showing significant deviation in PN/MOS2 flux in 2017. We assume that XMM-Newton and {\srg} captured {\srge} in the quiescent state in 2016 and 2019, respectively. 

\begin{figure}
    \centering
    \includegraphics[width=\linewidth]{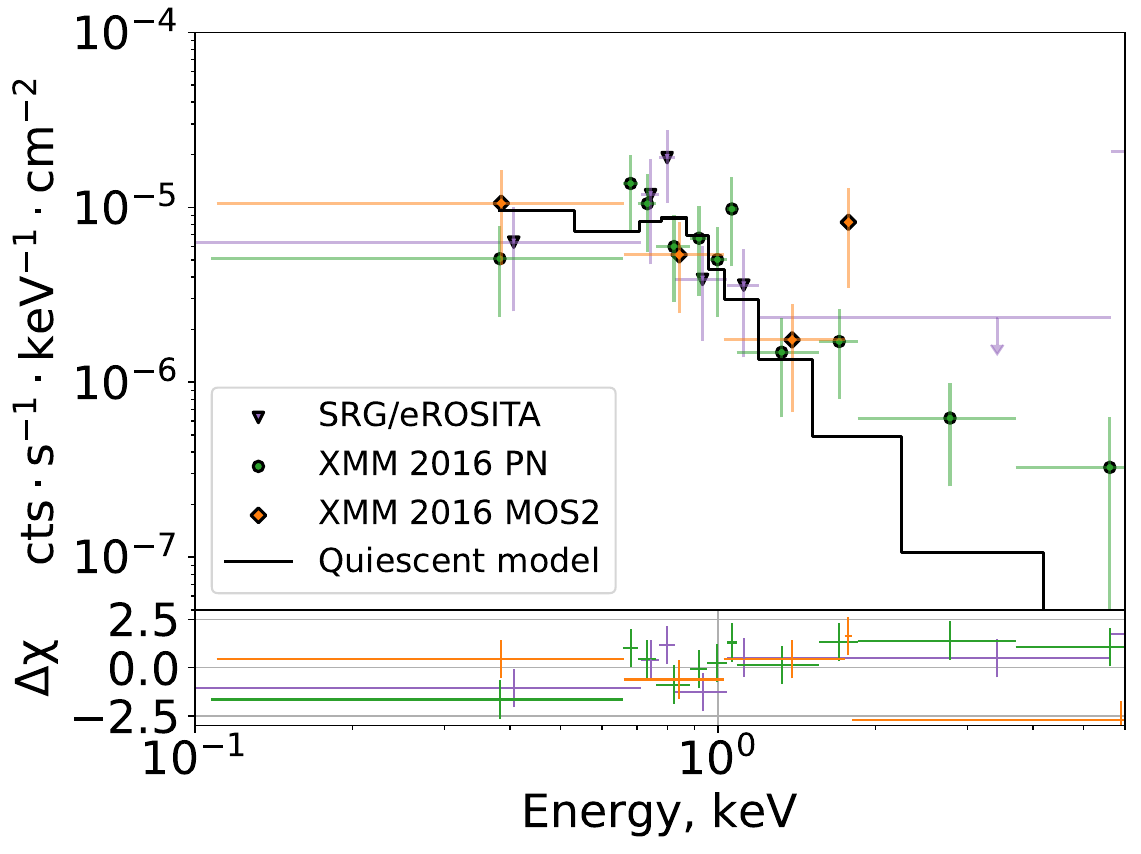}
    \caption{Quiescent X-ray spectrum of {\srge} based on the {\srg}/{\ero}, XMM 2016 PN and XMM 2016 MOS2 observations. Black solid line shows the best-fit model spectrum.}
    \label{fig:quiescent-spec}
\end{figure}

For the spectral analysis of the {\srge} in quiescent state, we combined 2016 and 2019 observations. Due to poor statistics, we assumed all three spectra to be described with a single model, consisting of the one-temperature plasma emission model \citep[$\tt APEC$,][]{Smith2001_APEC} modified by galactic hydrogen column density absorption ($\tt TBabs$). The galactic absorption is fixed to $\rm 1.91\times10^{20}~cm^{-2}$, the abundance of the plasma is set free and the redshift is fixed to zero. The best-fitting parameters with estimated 90\% confidence intervals are listed in Table~\ref{tab:spec_fit} and the spectrum with a model is shown in Figure~\ref{fig:quiescent-spec}. The temperature of the plasma in the quiescent state is estimated as ($0.8_{-0.2}^{+0.8}$~keV) or $\rm 9.3_{-2.3}^{+9.3} \times 10^{6}~K$. The emission measure in the quiescent state is estimated as $(2.4 \pm 1.0) \times 10^{52}~\rm cm^{-3}$. The abundance for the quiescent state is $0.06_{-0.03}^{+0.11}~Z_\odot$. Based on the estimated unabsorbed flux in 0.3$-$4.5 keV range of $(1.0 \pm 0.3) \times 10^{-14}~\rm erg~s^{-1}~cm^{-2}$ and the distance to {\srge} of $347 \pm 5~\rm pc$ we can estimate the unabsorbed X-ray luminosity in 0.3$-$4.5 keV range as $\rm L_{XQ} =(1.4 \pm 0.4) \times 10^{29}$~{\lum}.

\begin{figure}
    \centering
    \includegraphics[width=\linewidth]{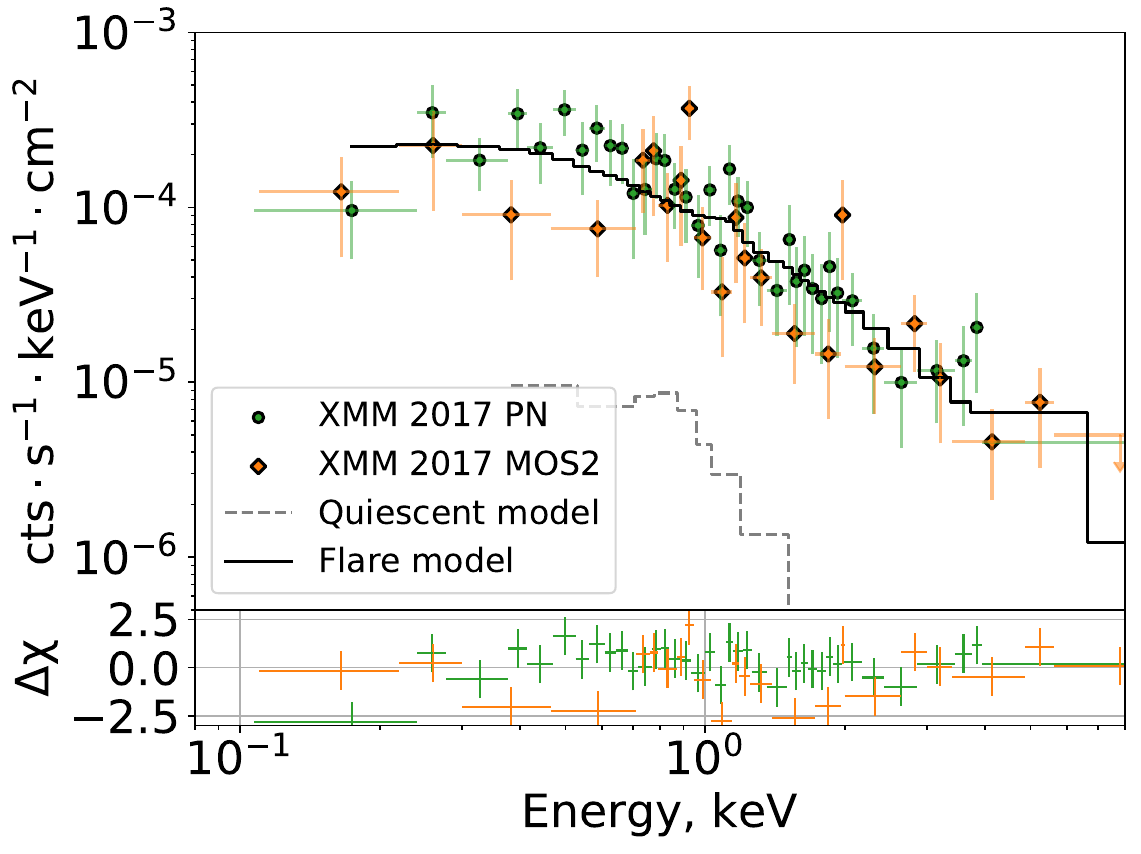}
    \caption{Flare X-ray spectrum of {\srge} based on the XMM 2017 PN and XMM 2017 MOS2 observations. Black solid line shows the best-fit flare model spectrum. Dashed grey line shows the quiescent model for comparison.}
    \label{fig:flare-spec}
\end{figure}

Similarly, PN/MOS2 flare spectrum in 2017 was fitted with one-temperature plasma model with fixed galactic absorption and free abundance (Table~\ref{tab:spec_fit} and Figure~\ref{fig:flare-spec}). The spectrum is characterized by the hot plasma component with the estimated temperature of $\rm 3.4_{-0.9}^{+1.4}~keV$ (or $\rm 3.9_{-1.0}^{+1.6} \times 10^7~K$) and emission measure of $\rm (5.2 \pm 1.0) \times 10^{53}~cm^{-3}$. The abundance during the flare was only estimated with the upper limit as the value $<0.82~Z_{\odot}$. The estimated unabsorbed luminosity of the flare is estimated as $\rm L_{XF} = (5.5 \pm 0.6) \times 10^{30}$ {\lum}. The total energy emitted during the flare in 0.3$-$4.5~keV, assuming the constant temperature during the flare, can be estimated as $\rm E_X = (1.7 \pm 0.2) \times 10^{34}~erg$.

\begin{table}
\begin{center}
\begin{tabular}{lll}
                               & Quiescent state               & Flaring state                 \\
\hline
$\rm N_H,~cm^{-2}$             & $1.91 \times 10^{20}$ (fixed) & $1.91 \times 10^{20}$ (fixed) \\
$kT,~{\rm keV}$                & $0.8_{-0.2}^{+0.8}$        & $3.4_{-0.9}^{+1.4}$           \\
$\rm EM, ~cm^{3}$              & $(2.4 \pm 1.0) \times 10^{52}$  & $(5.2 \pm 1.0) \times 10^{53}$  \\
$Z, ~Z_{\odot}$                & $0.06_{-0.03}^{+0.11}$        & $< 0.82$                      \\
Flux, $\rm erg~s^{-1}~cm^{-2}$ & $(1.0 \pm 0.3) \times 10^{-14}$ & $(3.8 \pm 0.4) \times 10^{-13}$ \\
cstat / dof                    & 102.7 / 91                    & 86.3 / 73
\end{tabular}
\end{center}
\caption{Results of spectral fitting using the single-temperature plasma models with fixed galactic column density for the spectra in quiescent and flaring states. The values are shown with 90\% confidence intervals. The unabsorbed flux is calculated in 0.3$-$4.5 keV range.}
\label{tab:spec_fit}
\end{table}

\section{Discussion}

The spectral analysis of {\srge} in the quiescent state revealed new indirect estimates for the stellar parameters. Both the emission measure and temperature in the quiescent state are in agreement with the values reported in \cite{Takasao2020_Gdwarfs} from the sample of 10 X-ray active G-dwarfs, however, the estimated quiescent luminosity of our star lies somewhat in between the two sample subdivisions to X-ray bright G-dwarfs and other Solar-type stars.

The X-ray luminosity of the {\srge} in quiescent state is $\sim$2 orders of magnitude higher compared to the Sun, which can be expected from the relatively young (0.6 Gyr) solar-type stars \citep{Gudel1997_Sun_in_time}, such as, for example, $\rm 9~Ceti~(BE~Cet)$ (Ra=00:22:51.7884553555, Dec=-12:12:33.969921319, J2000). Thus, the {\srge} is likely a young active star with age of $<$ 1 Gyr.

For direct comparison with the {\srg}/{\ero} all sky survey catalog, we recalculated the X-ray luminosity in the quiescent state in 0.3$-$2.3 keV range. The resulting value of $1.4_{-0.3}^{+0.5} \times 10^{29}~erg~s^{-1}$ is in agreement with the luminosity based on the 4.4 all sky surveys ($2.2 \pm 0.8 \times 10^{29}\rm~erg~s^{-1}$). Thus, two independent {\srg}/{\ero} observations of the {\srge} show no significant luminosity variations or Sun-like cycle activity across the 3-year span.

The estimated fluxes in the quiescent and flaring states show the $\sim 38$ times increase in flux. In this work, we assume that the estimation of the flux in the flaring state describes the flux of the flare itself, neglecting the contribution of the stellar flux of the {\srge} from outside the active region. In case the flux from outside the active region can be described by the flux in the quiescent state, it contributes less than 3\% of the observed flux during the flaring state. Thus, we assume that the spectrum, flux and luminosity estimations can be interpreted as the parameters dominated by the flare itself. 

The total X-ray energy emission $\rm E_X = (1.7 \pm 0.2) \times 10^{34}~erg$ during the flare of {\srge} exceeds $\rm 10^{33}~erg$, the canonical threshold for the superflares classification \citep{Schaefer2000_superflares}. Moreover, the simultaneous observations of superflares in optical and X-ray ranges \citep{Guarcello2019_Pleiades} show that the total emitted energy in optical range is similar or even exceeds the total energy emitted in X-rays, so the bolometric energy emission during the flare of {\srge} can be expected to be several times bigger than observed in X-rays. Thus, we confirm the classification of the observed event as a superflare on G-type star. The parameters of the superflare are in agreement with the analysis of a sample of X-ray stellar flares on cool (F-M spectral class) stars \citep{Pye2015_flare_survey}.

\subsection{Abundance estimation}
Abundance in the quiescent state is significantly sub-Solar but is in agreement with the abundances found from the low-resolution spectral analysis of the sample of X-ray active G-dwarfs \citep{Takasao2020_Gdwarfs}. The abundance found from the spectral modeling in this work is significantly lower than expected from the metallicity estimations from {\gaia} DR3, TIC-8 or APOGEE-2 SDSS DR17. It is likely related to the simplicity of the spectral model used in this work: as noted in \cite{Gudel2004_overview}, the low-resolution X-ray spectra are better described by the 2-temperature plasma models (for example, $\rm \tt apec + apec$ XSPEC model) or models with continuous power law emission measure-temperature distribution (for example, $\rm \tt cempow$ XSPEC model). In our case, however, the statistics did not allow for accurate parameter estimations with more complicated models. For example, fitting the quiescent spectrum using the $\rm \tt cempow$ with power law fixed at 1.0 incline results in the abundance estimation of $Z/Z_\odot = 0.3_{-0.2}^{+2.8}$.

The metallicities reported by the {\gaia} DR3, TIC-8 or APOGEE-2 SDSS DR17 are deviated from each other with the biggest outlier being the {\gaia} DR3 with significantly sub-Solar $[M/H] = -0.232 \pm 0.007$. This value contradicts the hypothesis of the {\srge} being a young solar-type star, as low element abundances are characteristic for the older stellar population. TIC-8 and APOGEE-2 SDSS DR17, on the contrary, show almost Solar element abundances, which is in agreement with the young star hypothesis.

\subsection{Estimation of the rotation period}

Quasi-periodic brightness modulations of the stars with observed superflares \citep{Maehara2012_superflares, Maehara2015_superflares, Maehara2017_starspots_flares} indicate the possible relation of superflare probability to the existence of huge starspots on the stellar surface. The lifetime of the starspots on solar-type stars ranges from several days to about a year \citep{Giles2017_lifetime, Namekata2019_lifetime}.

We searched for this variability using the open archive of Zwicky Transient Facility \citep{Bellm2019_ZTF} observations of {\srge} with data in the g filter from July 2018 to November 2023. We analyzed the periodicity in the data using Lomb-Scargle periodogram \citep{Lomb1976, Scargle1982}, but did not find any significant variability.

The source {\srge} was also in view by the TESS orbiting observatory \citep{Ricker2015_TESS} during two independent observing periods.

The first is from 19.10.2018 to 14.11.2018, which fell on sector 4 with a time resolution of 1800 seconds, and the second is from 22.10.2022 to 18.11.2022 corresponding to sector 31 with a resolution of 600 seconds. The duration of observations was $\sim25-27$ days for each sector. 

To search for the periodic variability of the source, the light curves available in the public domain on the MAST portal were used. Both light curves contain areas of discontinuity in the data of about 3 to 6 days duration. Therefore, the Lomb-Scargle method, which is well suited for analyzing non-uniformly distributed time series, was used to determine the period. The data analyses used PDCSAP\_FLUX fluxes, which are simple aperture photometry (SAP\_FLUX) values taking into account common instrumental effects of TESS detectors \citep{Ricker2015_TESS}.

\begin{figure*}
    \centering
    \includegraphics[width=\linewidth]{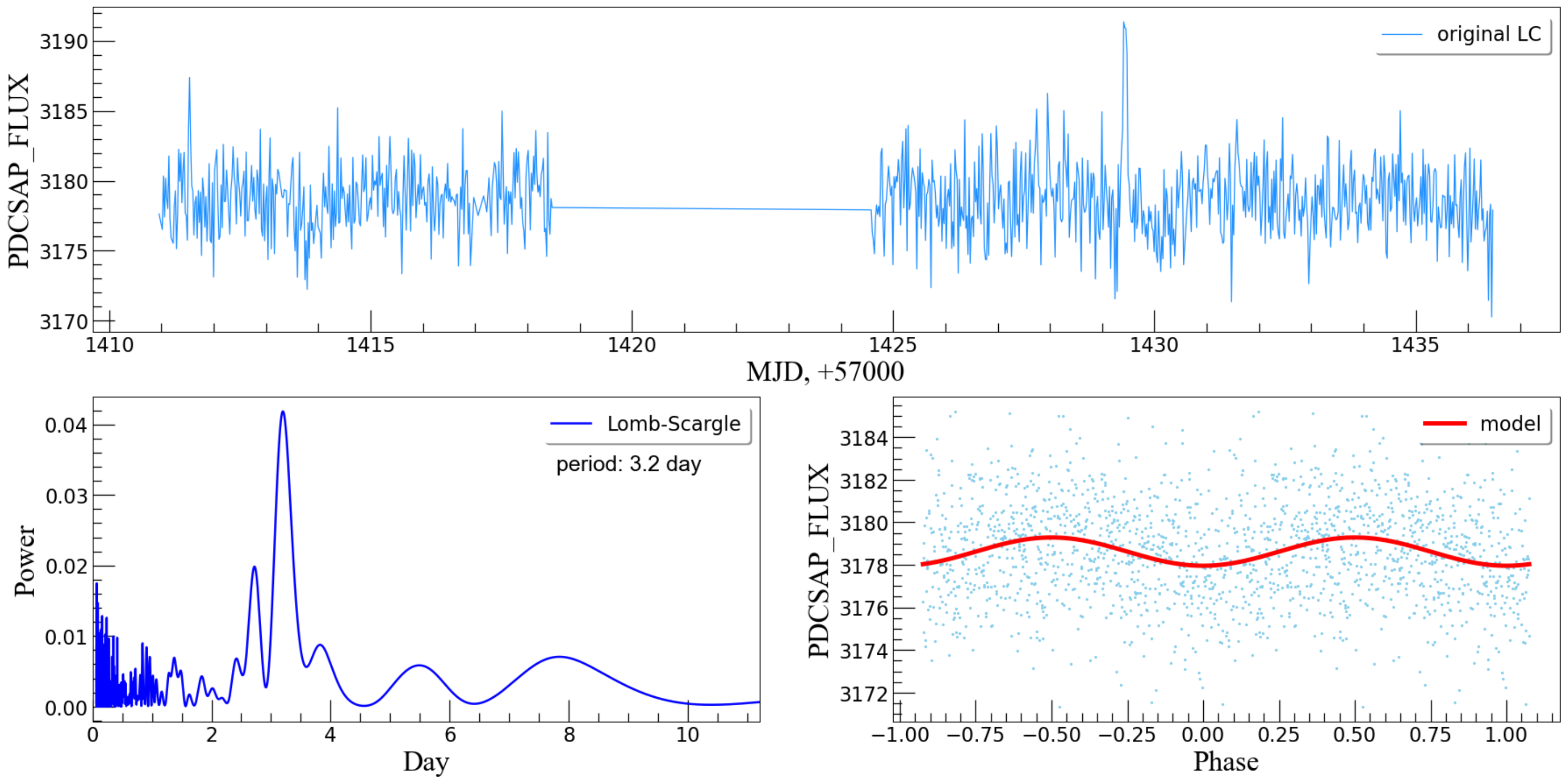}
    \caption{The top panel shows the {\srge} light curve for the sector 4 data. The bottom panel shows the periodogram and phase light curve. Red colour -- Lomb-Scargle model corresponding to the period of 3.2 days.}
    \label{fig:TESS_LC}
\end{figure*}

As a result, a period of $3.2 \pm 0.5$ days was obtained for two independent photometric series (Fig.~\ref{fig:TESS_LC}). The estimation of period accuracy is obtained from the half-width of the maximum peak on the periodogram. It is worth noting that with a photometric single measurement error of $\sim0.06\%$, the scatter in the light curve is only about $0.3\%$. When combining data for two sectors, the period value remains unchanged within the error limits. However, it is worth considering that the photometric noise in the light curve is much higher in the 31 sector data obtained at 600-second time resolution.

\subsection{Star spot area estimation} 

In addition to the search of rotational period of the star, we estimated the upper limit of the starspot area based on the amplitude of variations present in the data for both the ZTF and TESS light curves. We estimate the amplitude of variations as the difference between $95\%$-quartile borders observed in the apparent magnitude data, resulting in the upper limit in the amplitude of the flux variations to be $\rm \Delta F/F < 5.7~\%$. Following the estimation presented in the \cite{Maehara2017_starspots_flares}, we can calculate the relation of starspot area $\rm A_{spot}$ normalized by the area of solar hemisphere $\rm A_{1/2\odot} = 3 \times 10^{32}~\rm cm^{2}$ as:
\begin{equation}
    \rm
    \frac{A_{spot}}{A_{1/2\odot}} = (\frac{R_{star}}{R_{\odot}})^2 \times \frac{T_{star}^4}{T_{star}^4 - T_{spot}^4} \times \frac{\Delta F}{F}
\label{eq:starspot}
\end{equation}

where $\rm R_{star}$ and $\rm R_\odot$ are radii of the star and the Sun, respectively, $\rm T_{star}$ is an effective temperature of the star, $\rm T_{spot}$ is an effective temperature of the starspot, calculated as

\begin{equation}
    \rm T_{spot} = 0.751 \times T_{star} - 3.58 \times 10^{-5} \times T_{star}^{2} + 808 
\end{equation}

The resulting upper limit estimate for the starspot area of the star from Eq. \ref{eq:starspot} is $\rm A_{spot}/A_{1/2\odot} < 8\%$ for the ZTF data, which is in agreement with the observation of a large sample of G-stars with superflares from \citep{Maehara2017_starspots_flares}.

Based on the 0.3\% light-curve scatter of TESS data, we can estimate the upper limit on the observed starspot area as $\rm A_{spot}/A_{1/2\odot} < 0.4\%$. The observed normalized starspot area of $< 0.4\%$ is lower compared to the typical values reported in \cite{Maehara2017_starspots_flares} for stars with period of $\sim 3$ days, although there are examples of the superflares observed on stars with normalized starspot area of $\sim 0.1\%$. Despite that, further investigation of optical brightness modulation with lower uncertainties is required to estimate the starspot area.

\subsection{Luminosity-rotational period relation}
\begin{figure}
    \centering
    \includegraphics[width=\linewidth]{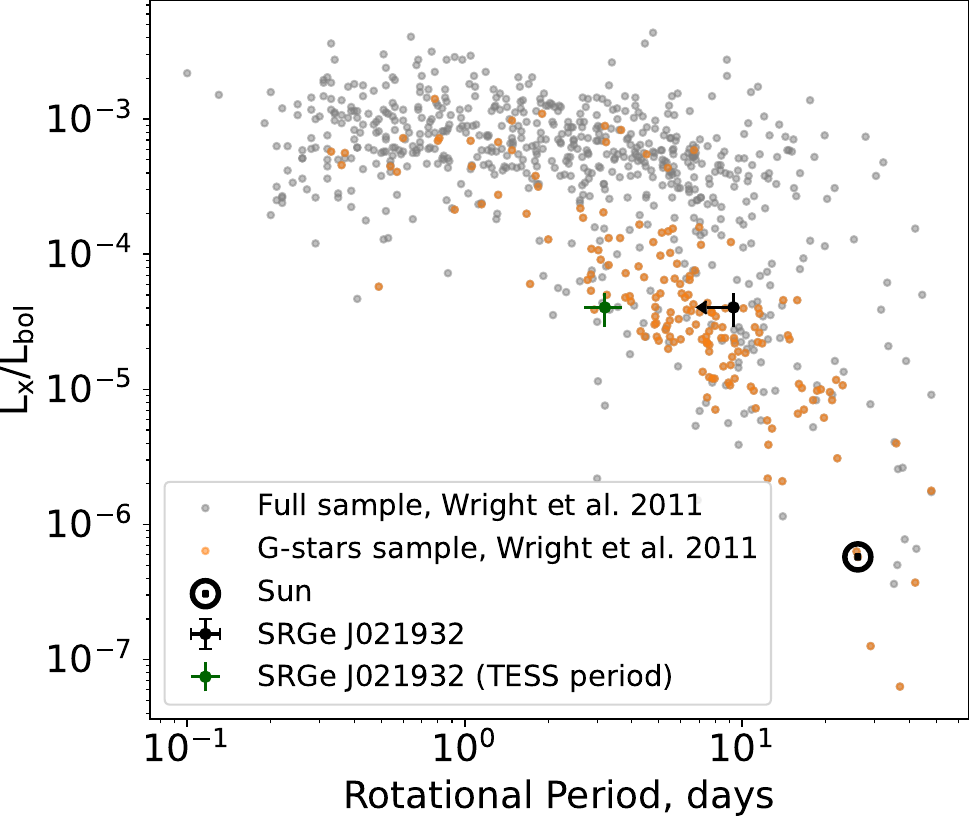}
    \caption{$\rm L_x/L_{bol}$ to Period relation for a sample of the stars from \citep{Wright2011_Lum_period_relation_new}. The estimated X-ray luminosity of {\srge} is recalculated to 0.1-2.4 keV to match the data of the sample. Grey dots show the full sample of the objects (824 F-M type stars), orange dots show the subsample of G-type stars (166 stars), black circle shows the Sun. The black cross shows the {\srge} in a quiescent state with the 90\% confidence interval errorbar and upper limit on the rotational period from APOGEE$-$2 data. The green cross shows the {\srge} with the period estimated from the TESS data.}
    \label{fig:L-P relation}
\end{figure}

Another possible source of information is the relation $\rm L_X / L_{bol}-P_{rot}$ shown in Fig.~\ref{fig:L-P relation} that can be used to determine, if the star is observed in X-ray saturated state \citep{Pizzolato2003_Lum_period_relation, Wright2011_Lum_period_relation_new}. The unabsorbed X-ray luminosity of the quiescent star is estimated as $\rm L_{XQ} = (1.5 \pm 0.4) \times 10^{29}$ {\lum} in 0.3$-$4.5~keV. For the direct comparison with the existing samples, we scaled the luminosity to the ROSAT typical energy range of 0.1$-$2.4 keV used in \citep{Wright2011_Lum_period_relation_new}, resulting in $\rm L_{X} = (1.8 \pm 0.5) \times 10^{29}$ {\lum}. The bolometric luminosity of the star from {\gaia} is $\rm L_{bol} = 1.19 \pm 0.04~L_{\odot}$. Thus, the relation between X-ray and bolometric luminosities is $\rm L_X / L_{bol} = (4.0 \pm 1.1) \times 10^{-5}$.

Based on the TESS light-curve analysis, we estimate the rotation period as $3.2 \pm 0.5$ days. Additionally, considering the $\rm v~sini = 6.1~km~s^{-1}$ based on the APOGEE$-$2 data and {\gaia} estimation for the radius $R = 1.04 \pm 0.01 R_\odot$, we can calculate the upper limit for the rotational period: $\rm P_{rot} < 9.3~days$.

The resulting position of the {\srge} on the $\rm L_X / L_{bol}-P_{rot}$ diagram is overplotted on Fig.~\ref{fig:L-P relation}. This position aligns well with both the given empirical relation for the full sample and the subsample of G-type stars. Additionally, it is evident that {\srge} was observed in a non-saturated state, that is describing the plateau at a $\rm L_{X} / L_{bol} \sim 10^{-3}$.

\subsection{Binary system hypothesis interpretation}

In the above sections, we mainly presented the interpretation of the results for single star hypothesis. As we mentioned in Section \ref{sec:binary}, however, there are at least two indicators for the radial velocity variations present in {\gaia} DR3 and APOGEE-2 SDSS DR17 catalog data. Thus, it is necessary to suggest an alternative interpretation in case we are working with the data from a binary system.

One possible solution to the observed data is the binary system of G-type and M-type dwarfs, in which case the optical and IR data might have not detected the M-type companion due to 1-2 orders of magnitude lower optical luminosity. The X-ray luminosity of the M-type companion, however, can be of the $\sim 10^{29}~\rm erg~s^{-1}$ \citep{Magaudda2020_Mdwarfs} which would be in agreement with the observed X-ray data. The superflaring activity then would also be associated with the M-type companion, which would satisfy the common trend stated by \cite{Karmakar2022_EQPeg}.

Rejection of either of the presented hypotheses requires additional regular observations to better determine the radial velocity variations, to search for possible eclipsing events, or to identify the companion using e.g. characteristic lines in IR spectra of the object.

\section{Conclusion}

In this work we studied the X-ray source {\srge}, that we associated with an X-ray active star of spectral class G2V$-$G4V and the rotational period $\rm P_{rot} < 9.3$ days.

{\srge} was detected by {\srg}/{\ero} during the eUDS X-ray survey in 2019. The observed flux, however, turned out to be much dimmer compared to that reported in {\xmm} catalogue 4XMM-DR12. Detailed analysis revealed that {\xmm} captured the source during a flaring event on January 1 2017.

The analysis of the {\xmm} light curve showed that the X-ray flare can be described by the Gaussian rise and exponential decay with characteristic scale  ${\sim}400$~s and ${\sim}1300$~s, respectively.

In the X-ray spectral analysis, aside from the flare event, {\srge} shows no outstanding features in its quiescent state with $\rm L_{XQ} = (1.4 \pm 0.4) \times 10^{29}$ {\lum} in 0.3$-$4.5 keV, well aligning with the sample of X-ray active G-dwarfs. The spectrum is well fitted with the one-temperature plasma emission model (APEC) with the estimated temperature of ${\sim}10$ MK.

The spectral analysis of the flaring state showed that the flare spectrum can also be described with the one-temperature plasma emission model (APEC) with the estimated temperature of ${\sim}40$~MK. The luminosity of the observed flare is $\rm L_{XF} = (5.5 \pm 0.6) \times 10^{30}$~\lum\ in 0.3$-$4.5 keV. The total energy emission during the flare exceeds the canonical threshold of $10^{33}$ erg, so we classify the observed event as a superflare on a solar-type star.

The estimated relation of X-ray luminosity to bolometric luminosity implies the {\srge} to be in a non-saturated state. The analysis of Zwicky Transient Facility data was used to calculate the upper limit on the starspot area to be $\rm A_{spot}/A_{1/2\odot} < 8\%$.

The analysis of TESS light-curves revealed the rotational period estimation of $3.2 \pm 0.5$ days and a tighter upper limit on the starspot area $\rm A_{spot}/A_{1/2\odot} < 0.4\%$, which is lower compared to the typical values reported in the previous observations of starspots on the solar-type stars with observed superflares.

Additionally, we report the peculiar features observed in the radial velocity estimations presented in APOGEE-2 and {\gaia} data, which can indicate the {\srge} to be a binary system. One possible interpretation of the observed superflare, in this case, would be that the binary system consists of the G-type and M-type stars, the latter of which remained undetected due to a lower optical luminosity while being responsible for the majority of X-ray flux and the superflaring activity. However, further observation of the source is required to reject either single G-star or binary system hypotheses.

\section*{Acknowledgements}

This work is based on observations with eROSITA telescope onboard SRG observatory. The SRG observatory was built by Roskosmos in the interests of the Russian Academy of Sciences represented by its Space Research Institute (IKI) in the framework of the Russian Federal Space Program, with the participation of the Deutsches Zentrum für Luft- und Raumfahrt (DLR). The SRG/eROSITA X-ray telescope was built by a consortium of German Institutes led by MPE, and supported by DLR. The SRG spacecraft was designed, built, launched and is operated by the Lavochkin Association and its subcontractors. The science data are downlinked via the Deep Space Network Antennae in Bear Lakes, Ussurijsk, and Baykonur, funded by Roskosmos. The eROSITA data used in this work were processed using the eSASS software system developed by the German eROSITA consortium and proprietary data reduction and analysis software developed by the Russian eROSITA Consortium. Based on observations obtained with XMM-Newton, an ESA science mission with instruments and contributions directly funded by ESA Member States and NASA. Based on observations obtained with the Zwicky Transient Facility project.
This paper includes data collected by the TESS missions and obtained from the MAST data archive at the Space Telescope Science Institute (STScI). Funding for the TESS mission is provided by the NASA Explorer Program.

The work of IB, MG, IKh  supported in part by the RSF grant N 23-12-00292 (source binarity analysis based on TESS photometry,  SDSS and GAIA spectroscopy data)

\section*{Data Availability}

{\xmm} data have been obtained through the XMM-Newton Science Archive. ZTF data have been obtained through the NASA IPAC Infrared Science Archive. TESS data have been obtained through the Mikulski Archive for Space Telescopes at the Space Telescope Science Institute (STScI). {\srg}/{\ero} spectrum published in this paper can be made available upon a reasonable request. 

\bibliographystyle{elsarticle-harv} 
\bibliography{refs}






\end{document}